\documentclass[aps,pra,twocolumn,superscriptaddress,showpacs,floatfix]{revtex4}
\usepackage[dvipdfmx]{graphicx,color}
\usepackage{color}
\usepackage{amsmath}

\begin{document}

\title{Quantum lock-in detection of a vector light shift}
\author{Kosuke Shibata}
\email{shibata@qo.phys.gakushuin.ac.jp}
\author{Naota Sekiguchi}
\author{Takuya Hirano}
\affiliation{Department of Physics, 
Gakushuin University, Tokyo, Japan}
\date{\today}

\begin{abstract}
We demonstrate detection of a vector light shift (VLS) using the quantum lock-in method.
The method offers precise and accurate VLS measurement without being affected by real magnetic field fluctuations.
We detect a VLS on a Bose--Einstein condensate (BEC) of $^{87}$Rb atoms 
caused by an optical trap beam with a resolution less than 1 Hz.
We also demonstrate elimination of a VLS by controlling the beam polarization
to realize a long coherence time of a transversally polarized $F$ = 2 BEC.
Quantum lock-in VLS detection should find wide application, 
including the study of spinor BECs, electric-dipole moment searches, and precise magnetometry.
\end{abstract}

\maketitle
\section{Introduction}
The a.c. Stark shift or \textit{light shift} plays significant roles in atomic physics.
One example is the optical trap \cite{Grimm2000}, 
which has been extensively used in cold atom experiments
and has been the subject of intriguing and important research, 
including low-dimensional \cite{Gorlitz2001} and uniform gases \cite{Gaunt2013}, and 
atoms in an optical lattice with applications to quantum simulation \cite{Bloch2008}
and atomic clocks \cite{Derevianko2011,Katori2011}.
It has also enabled the study of
multi-component gases and, in particular, spinor Bose--Einstein condensates (BECs) 
\cite{Stamper-Kurn2013}.

The light shift has vector and tensor components and hence is state-dependent in general \cite{Deutsch1998,Geremia2006,Deutsch2010}.
The state dependence has been exploited for realizing state-selective transport 
\cite{Mandel2003a, Mandel2003b} and confinement \cite{Heinz2020}.
However, a state-dependent shift is often undesirable for situations in which well-controlled spin evolution is required.
Escaping from a vector light shift (VLS), which is equivalent to a \textit{fictitious} magnetic field, has been an important issue in precise measurements, such as 
the search for an atomic electric-dipole moment \cite{Romalis1999} and exotic spin-dependent interactions \cite{Kimball2017PRD}.
Reducing the VLS is also important in atomic magnetometers,
in which the VLS introduces systematic errors. The quantum noise associated with the light shift due to the probe field
ultimately limits the sensitivity \cite{Fleischhauer2000}.

The VLS restricts the potential use of optically trapped atoms for magnetically sensitive experiments.
While its effect can be diminished by applying a bias magnetic field in a direction orthogonal to the wavevector,
the VLS can still be a significant noise source in precise measurements \cite{Romalis1999}.
It is necessary to reduce the VLS when an ultralow magnetic field is required.
In addition, the relative direction cannot be chosen satisfactorily
in some situations, such as in 3D optical lattice experiments.

In order to eliminate the VLS caused by optical trapping beams, the light polarization should be precisely controlled, because the VLS is proportional to the intensity of a circularly polarized component \cite{Grimm2000, Deutsch1998,Geremia2006,Deutsch2010}.
However, it is a formidable task to precisely extinguish the circular component at the atom position located in a vacuum cell.
Polarization measurements and control outside the cell do not assure the degree of linear polarization
due to the stress-induced birefringence of the vacuum windows \cite{Jellison1999}.

Therefore, a sensitive and robust polarization measurement method using atoms themselves as a probe is important. 
Most effective polarization measurements are accomplished by using atoms themselves as a probe.
Polarization measurements with an atomic gas have been performed 
with various methods including Larmor precession measurement \cite{Zhu2013}, precise microwave spectroscopy \cite{Steffen2013},
and frequency modulation nonlinear magneto-optical rotation \cite{Kimball2017}.
Differential Ramsey interferometry has been developed for spinor condensates \cite{Wood2016}.
Polarization measurements by fluorescence detection have been recently demonstrated for ions \cite{Yuan2019}.

In this paper, we demonstrate VLS detection by applying the quantum lock-in method \cite{Kotler2011, Lange2011}.
The measurement is immune to environmental magnetic field noise,
and thus achieves excellent precision and accuracy.
We detect a VLS induced by an optical trap beam on a BEC of $^{87}$Rb atoms with a resolution less than 1 Hz.
This detection method is feasible to implement
and should have wide applications in various research areas involved with optical fields.

The paper is organized as follows.
In Sec. \ref{sec: method}, our experimental method and setup are presented.
The experimental results are described in Sec. \ref{sec: result}.
We discuss the applications and potential performance of the quantum lock-in VLS detection in Sec. \ref{sec: discussion}.
We conclude the paper in Sec. \ref{sec: conclusion}.

\section{Experimental method and setup}\label{sec: method}
\begin{figure}
 \centering
 \includegraphics[width=8.5cm]{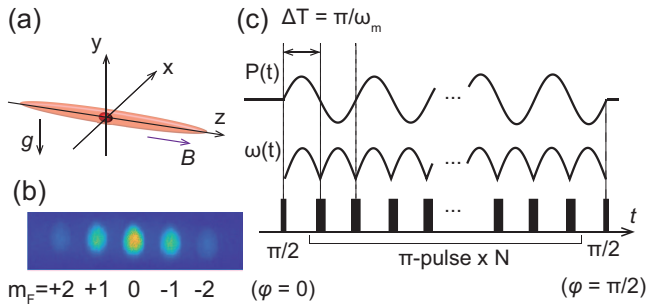}
 \caption{(color online) (a) Experimental configuration. 
A BEC is trapped in the axial trap beam along the $z$ axis and 
the radial trap beam along the $x$ axis (not shown).
(b) Typical TOF image of a BEC measured after rf pulses for the detection.
The spin components ($m_F=-2,-1,0,1,2$) are spatially resolved by the Stern--Gerlach method.
(c) Time sequence for the quantum lock-in detection of a VLS. 
The beam power, $P(t)$, is modulated with a frequency $\omega_m$.
The phase of the spin vector evolves with an angular frequency of $\omega(t) $.
The accumulated phase, $\Phi = \int_0^T \omega(t) dt$, is finally measured.
}
\label{fig1}
\end{figure}

We produce a BEC in a vacuum glass cell.
A BEC of $3 \times 10^5$ atoms in the hyperfine spin $F=2$ state is trapped in a crossed optical trap.
The trap consists of an axial beam at the wavelength of 850 nm and a radial beam at 1064 nm.
The axial and radial beam waists are $\approx$ 30 $\mu$m and 70 $\mu$m, respectively.
A magnetic bias field $B$ of 15 $\mu$T is applied along the axial beam to define the quantization axis, as shown in Fig.~\ref{fig1}(a).
The atoms are initially in the $|F,m_F \rangle = |2,2 \rangle$ state, where $m_F$ denotes the magnetic sublevel.
The ellipticity of the axial beam at the atomic position is controlled with a quarter waveplate (QWP) in the VLS measurement described below.
The QWP is located between a polarization beam splitter for polarization cleaning and the cell.
The angle of the QWP is adjusted with a precise manual rotation stage.
The minimum scale of the rotation stage is 0.28 mrad.

The time sequence for the quantum lock-in detection of a VLS is shown in Fig.~\ref{fig1}(c).
The lock-in technique enables enhanced sensitivity at the modulation frequency while reducing the effect of unwanted noise.
We measure a VLS induced by the axial optical trap beam with multiple rf pulses.
The trap beam power, $P(t)$, is modulated with a frequency $\omega_m$ during the pulse application as
\begin{equation}
P(t) = P_0\left( 1 + p \sin(\omega_m t) \right) \equiv P_0 + P_1 \sin(\omega_m t),
\end{equation}
where $P_0$ is the mean power and $p$ is the modulation index.
$P_1$ can be negative by changing the modulation phase by $\pi$.
$\omega_{m}$ is set to be sufficiently higher than twice the trapping frequency to avoid parametric heating of the atoms.
The modulation generates an a.c. fictitious magnetic field
to be measured, given by
\begin{equation}
B_{\mathrm{fic}} = -\frac{1}{4}\alpha^{(\mathrm{1})} \mathcal{C} I_{1} \sin(\omega_m t) \equiv B_{1} \sin(\omega_m t),
\end{equation}
where $\alpha^{(\mathrm{1})}$ is the a.c. vector polarizability, 
$\mathcal{C}$ is the degree of the circularity and $I_{1} $ is the beam intensity corresponding to $P_{1}$. 

The pulse set consists of an initial $\pi/2$ pulse at $t=0$, an odd number ($N$) of $\pi$-pulses, and a readout $\pi/2$ pulse.
The pulses are equally spaced by $\Delta T$.
The spacing satisfies $\omega_m$$=\pi/\Delta T$
so that the evolved phase due to the fictitious field is constructively accumulated.
The relative phase, $\Delta \varphi$, between the initial and read-out pulses 
is set to $\pi/2$ for maximum sensitivity to small changes in 
the accumulated phase, $\Phi$. $\Phi$ is explicitly given by
\begin{equation}\label{eq: phase}
\Phi = \frac{2}{\pi} \frac{g_F \mu_B B_1}{\hbar}T \equiv \frac{2}{\pi} \omega_{1} T,
\end{equation}
where $g_F$ is the Land\'{e} g-factor, $\mu_B$ is the Bohr magneton,
$\hbar$ is the reduced Planck constant and $T=(N+1)\Delta T$ is the phase accumulation time.
$\omega_{1}/(2\pi)$ represents the VLS corresponding to $B_1$ in units of frequency.

The read-out pulse converts $\Phi$ into the magnetization, $m$, as
\begin{equation}
m \equiv \frac{\sum_i i N_i}{N_{\mathrm{tot}}} = \mathcal{V} F \sin \Phi,
\end{equation}
where $N_i$ is the atom number in the $|F, m_F=i \rangle$ state $(i=-2,-1,0,1,2)$ after the read-out pulse, $N_{\mathrm{tot}}=\sum_i N_i$ is the total atom number, and $\mathcal{V}$ is the visibility.
$\mathcal{V}$ is ideally $1$, but in practice it is less than $1$
due to magnetic field noise \cite{Kotler2011}.
Imperfections in the initial state preparation and spin manipulation 
also decrease $\mathcal{V}$. 
The magnetization is measured by standard absorption imaging after a time-of-flight
with Stern--Gerlach spin separation (see Fig.~\ref{fig1}(b)).

\section{Results}\label{sec: result}
\begin{figure}
 \centering
 \includegraphics[width=8.5cm]{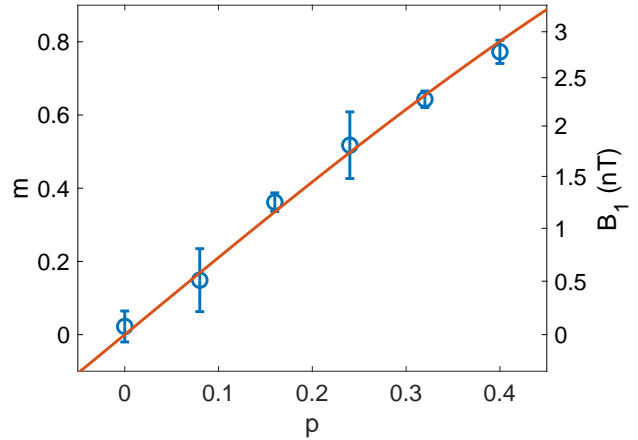}
 \caption{(color online) Detection of the VLS. 
The error bars represent the sample standard deviation.
The red solid line is the fitting curve by $\mathcal{V} F \sin(ap)$. 
The right axis represents $B_{1}$.
It should be noted that the right axis scale is not linear 
since $B_1$ is proportional to $\arcsin (\frac{m}{\mathcal{V}F}) $.
}
\label{fig: amplitude}
\end{figure}

We first confirm the validity of the detection scheme.
We perform a lock-in detection with $\omega_m = 2\pi \times 2$ kHz ($\Delta T = 0.25$ ms) and $N=27$, and hence $T = 7$ ms.
$P_0$ is fixed to 11 mW.
The change in $m$ is observed as $p$ is varied.
The result is plotted in Fig.~\ref{fig: amplitude}.
Here, the angle of the QWP axis, $\theta$, is approximately 4$^{\circ}$
apart from the optimal angle, $\theta^{*}$,
minimizing the VLS.
The experimental determination of $\theta^{*}$ is described below.
$m$ is well fitted by a sinusoidal function $\mathcal{V} F \sin(ap)$,
indicating the VLS was successfully detected.
The visibility in this detection setting is found to be $\mathcal{V} = 0.746 (42)$
from an independent measurement with no modulation ($p=0$) where $\Delta\varphi$ is scanned.

\begin{figure}
 \centering
 \includegraphics[width=8.5cm]{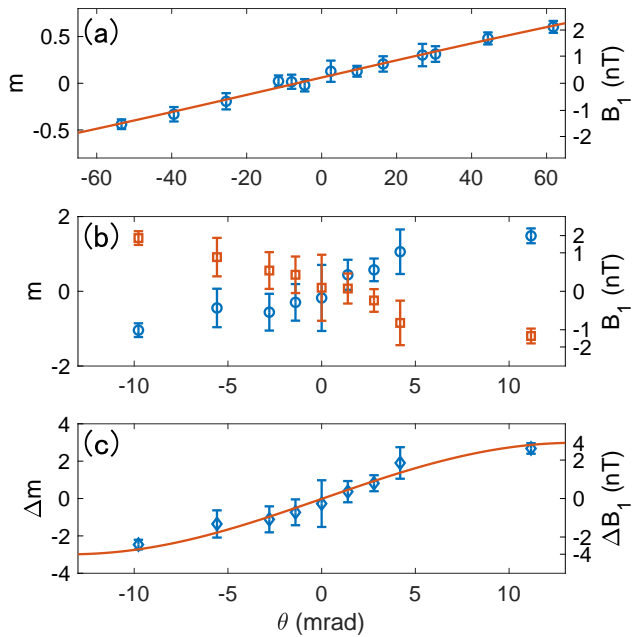}
 \caption{(color online) Polarization dependence of the signal.
(a) Measurement result with $T$ = 7 ms. 
(b) Measurement result with $T$ = 28.2 ms. 
The blue circles and red squares represent $m_+$ and $m_-$, respectively.
(c) $\Delta m$ as a function of $\theta$. 
$\Delta B_1$ is the difference between the fictitious magnetic fields 
for positive and negative $P_1$.
The solid lines in (a) and (c) are the fitting curves.
}
\label{fig: angle}
\end{figure}

The detection is used to minimize the VLS.
We control the VLS by changing $\theta$ with $p$ fixed to 0.32. 
The $\theta$-dependence of $m$ is shown in Fig.~\ref{fig: angle}(a).
Because $\mathcal{C} \approx \sin 2(\theta-\theta^{*}) \equiv \sin2\Delta \theta$
when the birefringence in the optical path is small \cite{Wood2016}
and $|\Delta \theta| \ll 1$,
we fit $m$ by $\mathcal{V} F \sin(\beta_1(\theta-\theta^{*}))$.
The fit gives $\beta_1 = 6.2(2)$, which is in reasonable agreement with the calculation.
$\theta^{*}$ is found to be -6.6(1.1) mrad.
The VLS resolution is evaluated as $\delta \omega = \beta_1 \delta\theta^{*}/T
= 2\pi \times 0.16$ Hz, where $\delta\theta^{*}$ is the uncertainty in the 
$\theta^{*}$ estimation.

We perform a fine estimation of $\theta^{*}$ 
by extending $T$ to 27.2 ms and applying a larger modulation.
In this experiment, $\omega_m$ and $N$ are $2\pi \times 625$ Hz and $33$, respectively.
We measure $m$ for $P_1 = \pm 13$ mW, referred to as $m_{\pm}$, respectively.
In finding $\theta^{*}$, we use $\Delta m = m_+ - m_-$
to cancel the offset due to the background field 
and the systematic error in the spin measurement.
The results are shown in Figs.~\ref{fig: angle}(b) and (c).
$\Delta m$ is fitted by $4\mathcal{V} F \sin(\beta_2 (\theta - \theta^{*}))$,
giving $\beta_2 = 117 (16)$ and $\theta^{*} = 0.06$ mrad 
with $\delta\theta^{*}$ = 0.40 mrad.
The angle resolution is improved 2.8 times.

We observe a larger variance in $m$ in the experiments for the fine $\theta^{*}$ estimation.
The standard deviation of $\Delta m$ is on average 0.64, 
while that for the reference data without modulation is $0.09$.
Therefore, a further improvement by a factor of at least 7 is possible,
because $\Delta m$ should ideally be independent of $T$ and the modulation strength.
We ascribe the increased variance to the actual variation of the vector shift over the experimental runs, caused by beam polarization fluctuation.
The result of the sensitive detection implies that 
the beam circularity varies with the standard deviation of approximately $3 \times 10^{-3}$.
On the other hand, from an independent experiment,
we expect that the retardance of the QWP should vary by several mrad due to the temperature change in our experimental room 
(within $\approx$ 0.6 K with a period of around 20 minutes).

\begin{figure*}
\centering
 \includegraphics[width=1\linewidth]{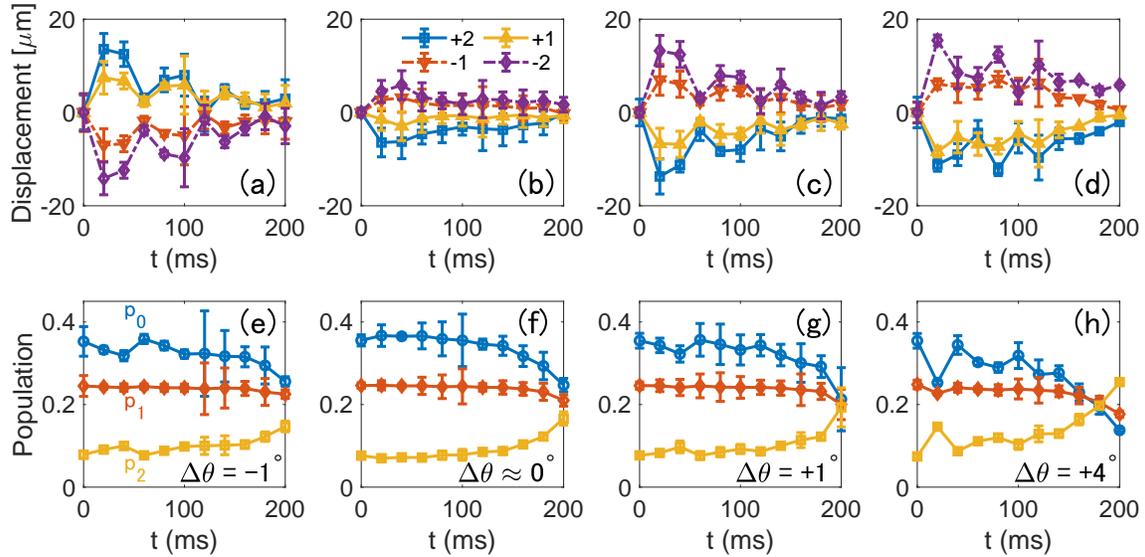}
 \caption{(color online) Effects of the fictitious magnetic field on a transversally polarized BEC. 
(a)--(d) Vertical displacement of the center of mass in the TOF image.
The panels show the data for $\Delta \theta$ = (-1, +0.02, +1, +4) degrees,
respectively. The solid and dashed lines are guides for the eyes.
(e)--(h) Population evolution corresponding to (a)--(d).
}
 \label{fig: position}
\end{figure*}

The BEC is subject to a fictitious magnetic field gradient without the VLS cancellation, because it is located at the shoulder of the optical trap beam due to gravity sag. 
While the observed fictitious magnetic field is small,
the gradient in the fictitious field can be on the order of 100 $\mu$T/m.
The gradient displaces the trap potential for each spin state other than the $m_F$ = $0$ state, thereby driving the spin dependent motion.
We observe an actual motion in a transversally-spin-polarized BEC in the hyperfine spin $F=2$ state, prepared after the initial $\pi/2$ pulse.
We plot the vertical displacement of the spin components in the TOF image,
which reflects the momentum, in Figs.~\ref{fig: position}(a)--(d).
The direction of the motion inverts depending on the sign of $\Delta \theta$
and the motion becomes small at $\Delta \theta \approx 0$.
These observations indicate that the motion is induced by the fictitious magnetic field.

The fictitious magnetic field gradient also causes nonlinear spin evolution
and thus a population change, as does the real magnetic field gradient \cite{Eto2014}.
The initial polarized atomic spin state breaks due to the spin mixing seeded by the nonlinear spin evolution.
We show $p_0 = N_0/N_{\mathrm{tot}}$, $p_1 = (N_{-1} +N_{+1})/(2N_{\mathrm{tot}})$,
and  $p_2 = (N_{-2} +N_{+2})/(2N_{\mathrm{tot}})$ in Figs.~\ref{fig: position}(e)--(h).
The population changes are observed at an earlier time ($t< 100$ ms) except for the case $\Delta \theta \approx 0$.
These changes can be attributed to the fictitious magnetic field gradient.
The faster population change for $\Delta \theta = 4^{\circ}$ is consistent 
with a qualitative estimation of the characteristic time for the change of 
$t_* \propto b^{-2/3}$, where $b$ is the magnetic field gradient \cite{Eto2014}.
A slow population change, which occurs regardless of $\Delta \theta$, 
is caused by a residual axial magnetic field gradient, $\partial B_z/\partial z$.
The existence of the axial gradient in these data is confirmed 
by the fact that the spin components separate in the axial direction at later times.

\begin{figure}
 \centering
 \includegraphics[width=8.5cm]{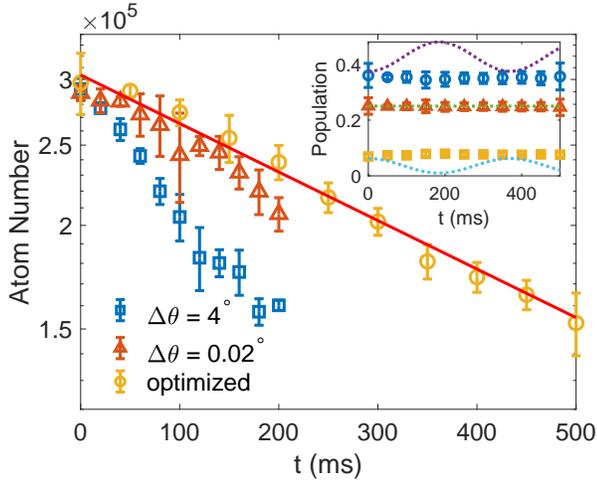}
 \caption{(color online) Atom number losses. The solid line is an exponential fit 
to the data with the optimized field gradient. 
The inset shows the population evolution for the optimized case.
The dotted lines are the prediction curves of the mean-field driven evolution without including the inelastic losses \cite{Kronjager2008}.
}
\label{fig: atomnumber}
\end{figure}

We next observed the change in the atom loss rate.
Figure ~\ref{fig: atomnumber} shows the evolution of the atom numbers, 
corresponding to the data in Figs.~\ref{fig: position}(b) and (d).
The decay is faster when $\Delta = 4^{\circ}$ than for $\Delta \approx 0^{\circ}$.
In the latter case, the decay rate starts to increase 
from around $t$ = 150 ms, where the population changes occur (see Fig.~\ref{fig: position}(f)). 
No increase in the loss rate is observed at the later time when we optimize $\theta$ and reduce the axial \textit{real} magnetic field gradient.
The $1/e$ time for the optimized condition is found to be 742 (31) ms.
The change of the loss rate can be understood from the property of the inelastic collisions 
in the $F=2$ state \cite{Tojo2009}.
Note that the inelastic collisional loss in the polarized state is inhibited  
due to the restriction of the angular momentum conservation.
The break of the polarized state due to the field gradient results in rapid atom losses.

However, the loss still occurs for the optimized condition.
Although the remaining loss may be due to the residual field inhomogeneity,
it is associated with the spin mixing 
induced by the quadratic Zeeman energy \cite{Kronjager2006, Kronjager2008}.
The model presented in \cite{Kronjager2006, Kronjager2008}, however,
needs to be modified to explain the observed population conservation,
shown in the inset of Fig.\ref{fig: atomnumber}.
According to \cite{Kronjager2008}, 
the population evolution in the limit of small quadratic Zeeman energy, $q$, is approximately given by
\begin{align}
p_0 =& \frac{3}{8}\left[ 1 + \frac{q}{2g_1n} (1- \cos(4g_1 nt/\hbar)) \right],\\
p_1 =& \frac{1}{4},\\
p_2 =& \frac{1}{16}\left[ 1 - \frac{3q}{2g_1n} (1- \cos(4g_1 nt/\hbar))\right],
\end{align}
where $g_1 = \frac{4\pi\hbar^2}{m}\frac{a_4-a_2}{7}$ is the interaction strength with $a_{\mathcal{F}}$ being the $s$-wave scattering length for the collisional channel of the total angular momentum $\mathcal{F}$ and $n$ is the mean atomic density.
Following these equations, $p_0$ and $p_2$ would undergo oscillations,
which is not in agreement with the observed experimental result.
We therefore attribute the population conservation to polarization purification
by inelastic collisional losses \cite{Eto2019}.
It should be noted that the observed population conservation contrasts with the case of the $F=1$ state, in which the magnitude of the polarization modulates \cite{Jasperse2017}.

\section{Discussion}\label{sec: discussion}
The quantum lock-in VLS detection is of practical use in cold atom experiments.
It can be used for evaluating the degree of circular polarization of an optical trap beam at the atomic position, as we have shown.
As the vacuum window birefringence introduces a maximum ellipticity 
of $10^{-2}$ or $10^{-1}$ \cite{Steffen2013},
a beam with no special care taken with respect to the $\textit{in vacuo}$ polarization 
may generate a fictitious field of several nT or a VLS of tens of Hz,
even with a shallow trap for ultracold atom experiments.
Quantum lock-in detection is sensitive enough to ensure better linear polarization at the atomic position
and therefore will greatly improve the magnetic conditions in cold atom experiments.
The sensitivity is sufficient to suppress the VLS below
the requirements for magnetically sensitive experiments, including studies of spinor BECs.
Although a homogeneous linear Zeeman shift does not 
affect the spinor physics due to spin conservation \cite{SKK2001},
a magnetic field gradient below several $\mu$T/m is typically required
to prevent magnetic polarization and observe the intrinsic magnetic ground state \cite{Stenger1998} or dynamics.
The sub-Hz VLS resolution of quantum lock-in detection meets this challenging demand.

Reducing the VLS is also important for precise measurements.
In addition to a direct energy shift, an inhomogeneous fictitious field is also detrimental to measurement accuracy \cite{Cates1988}.  
VLS reduction leads to a long coherence time, 
which is a mandatory requirement for highly sensitive measurements.
We have constructed a precise BEC magnetometer
using a transversally polarized $F=2$ BEC with a long coherence time,
realized using VLS elimination as we have shown.
The detail of the $F=2$ BEC magnetometer will be presented elsewhere \cite{Sekiguchi2020}.

We finally discuss the sensitivity limitations.
The sensitivity of the quantum lock-in detection is essentially the same 
as that of a Ramsey interferometer
with an equal phase accumulation time.
As the atom shot noise is dominant over the photon shot noise in typical absorption imaging,
the standard quantum limit in the VLS measurement is given by \cite{Giovannetti2004,Giovannetti2006}
\begin{equation}\label{eq: SQL}
\delta \omega = \frac{1}{T\sqrt{N_{\mathrm{tot}} } }.
\end{equation}
Here we replace the factor $\frac{2}{\pi}$ in Eq.~(\ref{eq: phase}) due to the sinusoidal
modulation with the maximal value of 1,
which is realized with a rectangular waveform modulation.
Substituting $N_{\mathrm{tot}} = 3 \times 10^5$ and $T$ = $30$ ms into Eq.~(\ref{eq: SQL}),
we obtain $\delta \omega = 2\pi \times 10$ mHz.
This is equivalent to a single shot field sensitivity of $\sim$ 1 pT.

\section{Conclusions}\label{sec: conclusion}
We have demonstrated precise detection of a VLS due to an optical trap
using the quantum lock-in method.
We have applied the detection to eliminating the VLS,
to observe the extension of the lifetime of transversally polarized
$F=2$ BEC.
The attained resolution of sub Hz is sufficient to suppress the VLS
below the required level for magnetically sensitive research, 
including the study of spinor BECs.
Although our demonstration was performed with a BEC,
the scope of the detection method is not limited to cold atom gases;
the proposed method can be applied to spin systems such as trapped ions and
diamond NV centers, where coherent spin control is possible.

\begin{acknowledgements}
This work was supported by the MEXT Quantum Leap Flagship Program (MEXT Q-LEAP) Grant Number JPMXS0118070326
and JSPS KAKENHI Grant Number JP19K14635.
\end{acknowledgements}


\end{document}